# Enhancement of ferromagnetism by Co and Ni substitution in the perovskite LaBiMn$_2$O$_{6+\delta}$

Asish K. Kundu,[*,a] V. Pralong,[a] V. Caignaert,[a] C. N. R. Rao[b] and B. Raveau[a]

[a]*CRISMAT Laboratory, 6 boulevard Maréchal Juin, Cedex 4, Caen-14050, France*

[b]*Chemistry and Physics of Materials Unit, Jawaharlal Nehru Centre for Advanced Scientific Research, Bangalore-560064, India*

## Abstract

The substitution of cobalt and nickel for manganese in the perovskite manganate LaBiMn$_2$O$_{6+\delta}$ has been realized, leading to the perovskites LaBiMn$_{2-x}$M$_x$O$_{6+\delta}$, with M = Co, Ni and $0 \leq x \leq 2/3$. In contrast to the literature those phases are found to be orthorhombic with *Pnma* symmetry. More importantly, it is shown that ferromagnetism is enhanced, T$_C$ being increased from 80 K for the parent compound ($x$ = 0) to 97 K for Ni-phase, and to 130 K for the Co-phase. Moreover, a strong competition between ferromagnetism and a glassy-ferromagnetic state in the case of nickel or a spin-glass behaviour in the case of cobalt is observed. These phenomena are interpreted in the frame of a phase separation scenario, where the ferromagnetic Mn$^{4+}$/Ni$^{2+}$ and Mn$^{4+}$/Co$^{2+}$ interactions reinforce the Mn$^{3+}$/Mn$^{4+}$ interactions. These compounds are found to be insulating with a relatively large positive value of thermoelectric power.

*E-mail: asish.kundu@ensicaen.fr; Tel: +33-231-452911; Fax: +33-231- 951600*

# I. INTRODUCTION

Bismuth based perovskites have been recognized recently as materials of potential interest for their eventual multiferroic properties as shown for instance from the studies carried out on $BiFeO_3$[1-4] and $BiMnO_3$.[5,6] In these oxides, magnetism originates from super-exchange interactions between iron or manganese cations through oxygen and ferroelectricity is most probably linked to the lone pair cation $Bi^{3+}$ which induces structural distortions. Thus, the research of ferromagnetic insulators containing bismuth is of importance in order to generate new magnetoelectric properties. In this respect, the perovskite oxides of the type $La_{1-x}Bi_xMnO_{3+\delta}$[7,8] are attractive, since they remain ferromagnetic up to $x = 0.5$, with a ferromagnetic curie temperature ($T_C$) of 80 K i.e. close to that of $BiMnO_3$ ($T_C \sim 100$ K) and can be prepared at normal pressure compared to the latter. Thus, Troyanchuk et al[7] reported the coexistence of ferromagnetic (FM) and paramagnetic (PM) state in $La_{1-x}Bi_xMnO_{3+\delta}$ when $x \leq 0.5$ and a cluster type spin-glass behaviour when $x = 0.65$. For $x = 0.5$, a cluster-glass magnetic phase has been reported by Zhao et al[8] who described this property due to the presence of magnetic clusters, hence to lack of good ferromagnetism. Based on previous studies of the ferromagnetism induced by nickel and cobalt substitution in $LaMnO_3$[9-11] and bearing in mind the results obtained for double ordered perovskites $La_2MnCoO_6$[12-14], $La_2MnNiO_6$[15-18] and $Bi_2MnNiO_6$[19-21], we have explored the effect of Co and Ni substitutions in the perovskites $LaBiMn_{2-x}M_xO_{6+\delta}$ for M = Co or Ni, with substitution level up to $x = 2/3$. We show that in the case of cobalt, ferromagnetism is significantly enhanced $T_C$ reaching a value of 130 K, with a much higher coercive field of 0.58 Tesla (T) compared to 0.05 T for $x = 0$ (at T = 10 K). In contrast, for nickel, ferromagnetism is increased at a lesser degree, with maximum $T_C$ of 97 K and the material remaining a soft ferromagnet, similar to $x = 0$. These results are interpreted in the frame of phase separation, involving strong FM super-exchange interactions between $Mn^{4+}$ and $Co^{2+}$ ($Ni^{2+}$) species.[12-21]

## II. EXPERIMENTAL PROCEDURE

The solid-state samples were prepared by conventional sol-gel method. Stoichiometric amounts of metal nitrates were dissolved in distilled water and citric acid and ethylenediamine were added to the solution in the molar ratio. The mixture solutions were stirred at 333 K for few hours and evaporated at 373 K to form gel. The gel mixtures were dried at 423 K for 12h and the resulting powders decomposed in air at 633 K for 2-3h, then at 1073 K for 12h inside a furnace. The powder samples were ground thoroughly and pressed into rectangular bars, and finally sintered at 1173-1223 K in platinum crucible for 24h and cooled rapidly to room temperature by taking out the samples from the furnace.

Small parts of the sintered bars were taken and ground to form fine powder to record the X-ray diffraction (XRD) pattern, using a Philips X-Pert diffractometer employing Cu-Kα radiation. The phases were identified by performing Rietveld[22] analysis in the 2θ range of 5°-120° and the lattice parameters are listed in Table 1. Composition analysis was carried out by energy dispersive spectroscopy (EDS) analysis using a JEOL 200CX scanning electron microscope, equipped with a KEVEX analyzer and it confirms the cationic composition within instrumental limitations. The oxygen stoichiometry was determined by redox titrations with $K_2Cr_2O_7$ 0.1N in acidic media (20mL HCl 2N + 10mL $H_3PO_4$ 1N).

Other pieces of the bars were taken for magnetization, resistivity and thermopower measurements. A Quantum Design physical properties measurements system (PPMS) was used to investigate the magnetic properties of the samples. The temperature dependence of the zero-field-cooled (ZFC) and field-cooled (FC) magnetization was measured in different applied magnetic fields. Hysteresis loops (M-H) were recorded at different temperatures. The dynamics of the magnetic response was studied by ac-susceptibility measurements at different frequencies in the low temperature range. In the measurements of temperature dependence of the ZFC magnetization, the sample was cooled from 300 K to 10 K in zero-field, the field was applied at 10 K and magnetization recorded on re-heating the sample. In the FC measurements the sample was cooled (from 300 K) in the applied field to 10 K and the magnetization recorded on re-heating the sample, keeping the field applied.

The electron transport (resistivity and thermopower) measurements were carried out by a Quantum Design PPMS magnetometer, with a standard four-probe method and a home made sample holder for thermoelectric power measurements in the temperature range of 10-400 K. The electrodes on the sample were prepared by ultrasonic deposition method using indium metal.

## III. RESULTS AND DISCUSSION

### A. Structure and chemical analysis

The chemical analysis of all samples show that three of them exhibit an oxygen excess (Table 1) with respect to the ideal perovskite formula, in agreement with the results previously reported for $LaBiMn_2O_{6+\delta}$.[7,8] Such results, already observed for $LaMnO_{3+\delta}$,[23] correspond to a cationic deficiency in the perovskite structure $La_{1-\delta'}Bi_{1-\delta'}(Mn,M)_{2(1-\delta')}O_6$ as previously explained by many authors. Nevertheless, we note that this excess oxygen tends to disappear rapidly, as manganese is replaced by cobalt or nickel. The compound $LaBiMn_{4/3}Co_{2/3}O_{6.02}$ being almost stoichiometric, where as LaBiMn$_{4/3}$Ni$_{2/3}$O$_{5.96}$ is oxygen deficient ($\delta \approx -0.04$) as shown in Table 1.

The Bi-based samples LaBiMn$_{2-x}$M$_x$O$_{6+\delta}$, with M = Co/Ni, could be obtained as single-phase over a small range of substitution, $0 \leq x \leq 2/3$, without any traces of impurities as shown from their X-ray powder diffractograms (Fig. 1). Beyond $x = 2/3$, traces of impurities, involving Bi$_2$O$_3$ or other phases of Bi-Mn-M-O systems, prevented the phases LaBiMn(M)O$_{6+\delta}$, with M = Co or Ni, to be isolated, in contrast to La$_2$MnCoO$_{6-\delta}$[12-14] and La$_2$MnNiO$_{6+\delta}$.[15-18] Refinements carried out by Rietveld[22] method, show that all samples including LaBiMn$_2$O$_{6+\delta}$ ($x = 0$), can be indexed in an orthorhombic structure, with the *Pnma* space group (Table 1). Note that this behaviour is different from that of the previously reported one for LaBiMn$_2$O$_{6+\delta}$, which was indexed either as rhombohedral[7] or as tetragonal.[8] Such a difference may originate from the completely different synthesis conditions and oxygen stoichiometry.

**B. Magnetic behaviour of the parent compound LaBiMn$_2$O$_{6.20}$ revisited**

Fig. 2a shows the magnetization curves, M (T), investigated in an applied field of 100 Oe confirms a broad transition from a PM state to a FM state around 80 K (T$_C$), as previously reported in the literature.[7,8] The existence of ferromagnetism in LaBiMn$^{III}_{1.6}$Mn$^{IV}_{0.4}$O$_{6.20}$ is in perfect agreement with the possibility of FM interactions between Mn$^{3+}$ and Mn$^{4+}$ species, since the average oxidation state of manganese is close to 3.20. Nevertheless the nature of the FM phase is not clear, since it was described as a cluster-glass by Zhao et al[8] and as a weak ferromagnet or rather as a coexistence of PM and FM phases by Troyanchuk et al.[7] Therefore, we have studied this low temperature phase in details at different applied fields to understand the nature of FM ordering.

Using a smaller field of 20 Oe, the ZFC curve shows much more narrow peak (Fig. 2b) than in 100 Oe (Fig. 2a) and moreover the irreversibility temperature ($T_r < T_C$) obtained from the ZFC and FC magnetization curves is increased. In contrast, for a higher field of 1000 Oe (inset Fig. 2b), both the ZFC and FC curves merge down to low temperatures and the peak in ZFC curve disappears completely, suggesting a long-range FM ordering. Nevertheless, the lack of magnetic saturation below the transition temperature and the low-fields study reveal that the ferromagnetism cannot be long-range order. The large divergence between ZFC and FC magnetization at low fields can be explained in the scenario of phase separation of FM domains involving $Mn^{3+}$ and $Mn^{4+}$ states distributed in an antiferromagnetic (AFM) matrix. Thus, during ZFC measurement the spins of magnetic ions will freeze in random direction and the magnetic anisotropic energy is relatively large. Therefore, the low magnetic fields are not sufficient to align them in the direction of the applied field; as a result strong divergence appears. But, the higher field will overcome this anisotropic energy and re-orient the spins in the field directions; hence there will be no divergence between ZFC and FC magnetizations.

The M (H), curves studied at different temperatures (Fig. 3a), also confirm this short-range ordering. One indeed observes a small hysteresis loop at 10 K, with a remanent magnetization ($M_r$) value of ~ 1.0 $\mu_B$/f.u. and a coercive field ($H_C$) of ~ 0.05 T. The highest value of magnetic moment is only ~ 6 $\mu_B$/f.u., which is less than the spin-only value of Mn-ions. At low temperature, the relatively smaller value of $H_C$ signifies a typical soft ferromagnet and at temperatures higher than $T_C$ (T > 100 K) the M-H behaviour is linear, corresponding to a PM state. Another interesting feature at low temperature is the unsaturated behaviour of M-H curve even at higher fields, which is a characteristic feature of glassy-ferromagnetic system.[24,25] To confirm this assumption we have studied frequency variation magnetic measurements below Curie temperature ($T_C$). Fig. 3b shows the in-phase $\chi'(T)$ component of the ac-susceptibility measured at four different frequencies. The in-phase $\chi'(T)$ data is similar to the low field ZFC magnetization curve, with a sharp peak around 80 K. A distinct frequency-independent peak corresponding to FM ordering and below this temperature a weak frequency-dependence behaviour is noticed. The magnetic ac-susceptibility behaviour observed for $LaBiMn_2O_{6.20}$ is quite different from canonical spin-glass system[26,27] and is akin to that of glassy-ferromagnetic materials.[24,25]

We have also investigated the PM region to clarify the magnetic interaction at high temperatures. Therefore, we have plotted inverse of magnetic susceptibility data with the variation of temperature in the range of 50-300 K (inset of Fig. 2b). The data clearly follows Curie-Weiss behaviour (for T > 200 K) and a linear fit to the Curie-Weiss law yields a PM Weiss temperature ($\theta_p$) of 120 K and an effective magnetic moment ($\mu_{eff}$) of ~ 7.9 $\mu_B$/f.u. The obtained value of $\theta_p$ is higher than the $T_C$ value (80 K), and the positive value of $\theta_p$ also signifies the FM interactions in the high temperature region.

**C. Magnetic behaviour of the substituted phases $LaBiMn_{2-x}M_xO_{6+\delta}$ (M = Co, Ni)**

The substitution of cobalt for manganese in $LaBiMn_2O_{6+\delta}$ ($x$ = 0), results in a significant increase of the FM transition temperature as shown in Fig. 4. The FM transition temperatures were calculated from the minimum position of the (dM/dT) vs temperature curves (see inset Fig. 4). Temperature dependence ZFC and FC magnetization data for all the samples exhibit a clear PM-FM transition at low temperature in an applied field of 100 Oe and the highest $T_C \approx 130$ K, is obtained for a doping concentration of $x$ = 2/3. The magnetization value increases with increasing the substitution level and the magnetization behaviour is almost same for all samples.

There is a large divergence between ZFC and FC data below the transition temperature similar to the parent compound ($x = 0$) at low fields; moreover the magnitude is higher than the latter. The high temperature region (T > 200 K) follows the Curie-Weiss law and the corresponding Weiss temperatures are 135 K and 150 K for $x = 1/2$ and 2/3 respectively as expected. This indicates that with increasing the cobalt content the FM interactions increase inspite of the decrease in oxygen content. Such a phenomenon can easily be explained by the presence of $Co^{2+}$ ions, which interact with $Mn^{4+}$ ions through positive super-exchange interactions.[12-14] In Fig. 5, we have presented high-field (H = 1000 Oe) magnetization data for both samples. Unlike $LaBiMn_2O_{6.20}$, the Co-substituted samples show considerably large divergence between ZFC and FC magnetization at low temperatures. With increasing field, the cusp in ZFC data becomes broad and shifts toward lower temperatures (does not disappear) and the FC magnetization shows no sign of saturation in the low temperature range. Therefore, we have studied the field variation of magnetization, M(H), for both samples at low temperatures (see insets of Fig. 5), which also show unsaturated values of magnetization, even at higher applied field (up to 5 T). The coercive field and remnant magnetization values for $x = 1/2$ are 0.18 T and 3.4 $\mu_B$/f.u., whereas for $x = 2/3$, they are 0.58 T and 2.5 $\mu_B$/f.u. respectively. From these data, it is clear that with increase in cobalt content $x$, the materials become magnetically hard at low temperature (10 K). With increase in temperature up to 100 K (below $T_C$), the materials turn to soft ferromagnet like $LaBiMn_2O_{6.20}$ (Fig. 3a).

The higher values of coercive field with increasing substitution levels explain the magnetic anisotropy behaviour below the FM transition. Due to this, there is a large divergence between the ZFC and FC curves even at 1000 Oe for both samples. But, the unsaturated value of FC magnetization (even at higher fields) is difficult to understand in the scenario of only short-range FM ordering. These behaviours may be due to electronic phase separation at low temperature, where large FM domains are present inside an AFM matrix. At low temperature, there is strong competition between positive FM ($Mn^{3+}$-$Mn^{4+}$ and $Mn^{4+}$-$Co^{2+}$) and negative AFM ($Mn^{4+}$-$Mn^{4+}$ and $Co^{2+}$-$Co^{2+}$) interactions and the FM interactions dominate over AFM at higher fields. But, the contribution of AFM interactions is not negligible; hence the M-H behaviour is exhibiting an unsaturated nature like glassy-ferromagnetic materials.[24,25]

In Fig. 6a, we show the effect of Ni for Mn substitution in $LaBiMn_2O_{6+\delta}$ ($x = 0$), leading to a similar behaviour to Co-substitution. The FM $T_C$ increases to a value of 97 K for $x = 2/3$, which is much lower than Co-substituted compound (see Fig. 4). The low-field ZFC and FC behaviour is almost similar to the samples discussed in previous sections, but the magnetic moment value is lower in magnitude. We have also studied the magnetic behaviour at higher fields (1000 Oe), which shows (Fig. 6b) similar properties to $x = 0$ compound (except higher $T_C$), with very small divergence between ZFC and FC data, quite unlike Co-substituted samples. The field variation of magnetization also confirms this nature with a smaller value of coercive field, 0.003 T at 10 K (0.05T for $x = 0$) and the highest magnetic moment is only 3.5 $\mu_B$/f.u. This is the lowest value of moment obtained in all the series of investigations, although the magnetic interactions should be similar in nature, i.e. positive (FM) interactions between $Mn^{3+}$-$Mn^{4+}$ and $Mn^{4+}$-$Ni^{2+}$ as reported in literature.[15-18] The relatively smaller value of moment and lower value of $T_C$ ($\cong$ 97 K) for this sample, compared to the cobalt phase can be explained by its oxygen deficiency, which induces a smaller $Mn^{4+}$ content and creates disordering on the cationic sites. Further investigations are necessary to understand this particular behaviour.

Fig. 7, shows in-phase ac-susceptibility versus temperature plot for LaBiMn$_{4/3}$M$_{2/3}$O$_{6+\delta}$ with M = Co, Ni. Similar to the parent compound (see Fig. 3b), these two samples also follow the low-fields ZFC magnetization data. The ac-susceptibility of Co-substituted sample reveals a weak frequency-dependent peak at low temperature, corresponding to the FM ordering. The position of peak at 10 Hz in an applied ac field of 10 Oe occurs at 125 K, and shifts to higher temperature (130 K) with increasing frequencies (10 kHz). But, the Ni-substituted sample does not show any shift in the peak temperature with varying frequencies although there is a change in magnitude with increasing frequencies. Hence, the Co-substituted sample has a frequency-dependent maximum in χ' (Fig. 7a), while the Ni-substituted sample reveals a similar feature to LaBiMn$_2$O$_{6.20}$ (see Fig. 3b). These data for LaBiMn$_{4/3}$Co$_{2/3}$O$_{6.02}$, are consistent with materials behaving as spin-glass[26,27] like system albeit of large magnetization values (ZFC, FC and M-H), below the transition temperature. In contrast, the parent compound and LaBiMn$_{4/3}$Ni$_{2/3}$O$_{5.96}$ samples show a glassy-ferromagnetic behaviour as reported in the literature.[24,25]

**D. Electron transport properties**

Temperature dependence of the electrical resistivity (ρ) behaviour for all samples is shown in Fig. 8. With decreasing temperature the resistivity increases for all samples and the value is very high at low temperature, crossing instrument limitations below 100 K. The rapid change in the temperature coefficient of resistivity (dρ/dT) from room temperature to low temperatures signifies the insulating behaviour. We have also studied the magnetoresistance effect in an applied field of 7 T and noticed that there is very small change in resistivity at low temperature. Thus, none of these samples show insulator-metal transition corresponding to FM T$_C$ and the magnetoresistance remains very small. In the 100 – 400 K range, the temperature variation of resistivity (with magnetic field of 0 and 7 T) confirms the insulating phase although the materials are FM below room temperature, as revealed from the magnetization results discussed earlier.

We have studied the thermoelectric power measurements to have an idea about the type of charge carriers in the materials, which also estimate the substitution level depending upon carrier concentration. The measurements were carried out as a function of decreasing temperature. Since all samples are poor conductors, with a high value of resistivity below room temperature, the data became unreliable below 150 K due to the impedance limits of the instrument. The value of Seebeck coefficient (S) for $LaBiMn_2O_{6.20}$ at 300 K is around +133 µV/K, and it increases slowly with decrease in temperature. On the other hand, $LaBiMn_{4/3}Co_{2/3}O_{6.02}$ shows an intermediate magnitude with a positive value of S ≈ 81 µV/K at room temperature, which confirms the p-type polaronic conductivity or hole-like carriers in the materials. This type of behaviour is also reported in the literature[10,11] for doped manganate system. Therefore both samples show a decrease of p-type polaronic conductivity with decreasing temperature as expected from the resistivity behaviour.

## IV. CONCLUSIONS

The prime result from the current investigations of the $LaBiMn_{2-x}M_xO_{6+\delta}$ series describes the enhancement of ferromagnetism by cobalt and nickel substitution at the Mn-site, with an increase of $T_C$ up to 130 K, while the materials remain insulating below $T_C$. The second important point concerns the glassy ferromagnetic behaviour at low temperature, with a significant difference between the Co-substituted phases, which can be considered as spin-glasses, and the Ni-substituted phases, which can be described rather as glassy ferromagnets. In any case, these results can be explained in the frame of phase separation scenario, involving FM domains embedded in an AFM matrix. In case of the parent compound $LaBiMn_2O_{6.20}$, the FM domains result from FM interactions between $Mn^{3+}$ and $Mn^{4+}$ species ($\delta > 0$), whereas in the substituted phases ferromagnetism is reinforced by FM interactions between $Mn^{4+}$ and $Co^{2+}$ or $Ni^{2+}$ species. The antiferromagnetic matrix results from AFM interactions of the $Mn^{3+}/Mn^{3+}$ species, which is increased by $Mn^{4+}/Mn^{4+}$ and $M^{2+}/M^{2+}$ AFM interactions during substitution. Hence, there is a strong competition between FM and AFM interactions below the transition temperature. The latter is at the origin of spin-glass behaviour in the Co-substituted phase and of glassy-ferromagnetic state in the Ni-substituted phase. The oxygen deficiency observed for nickel compound may be at the origin of the different behaviour of the cobalt phases compared to the nickel ones.

## Acknowledgements

The authors thank Dr. S. Hebert for help during SQUID measurements and S. Rey for technical help. We gratefully acknowledge the CNRS and the Minister of Education and Research for financial support through their Research, Strategic and Scholarship programs, and the European Union for support through the network of excellence FAME.

**References:**

1 G. A. Smolenskii and I. E. Chupis, *Sov. Phys. Usp.*, 1982, **25**, 475.

2 I. Sosnowska, T. Peterlin-Neumaier and E. Steichele, *J. Phys. C*, 1982, **15,** 4835.

3 Y. Kuroiwa, S. Aoyagi, A. Sawada, J. Harada, E. Nishibori, M. Takata and M. Sakata, *Phys. Rev. Lett.*, 2001, **87**, 217601.

4 J. Wang, J. B. Neaton, H. Zheng, V. Nagarajan, S. B. Ogale, B. Liu, D. Viehland, V. Vaithyanathan, D. G. Schlom, U. V. Waghmare, N. A. Spaldin, K. M. Rabe, M. Wuttig and R. Ramesh, *Science*, 2003, **299,** 1719.

5 A. M. Santos, S. Parashar, A. R. Raju, Y. S. Zhao, A. K. Cheetham and C. N. R. Rao, *Solid State Comm.*, 2002, **122,** 49.

6 T. Kimura, S. Kawamoto, Y. Yamada, M. Azuma, M. Takano and Y. Tokura, *Phys. Rev. B*, 2003, **67**, 180401R.

7 I. O. Troyanchuk, O. S. Mantytskaja, H. Szymczak and M.Y. Shvedun, *Low. Temp. Phys.*, 2002, **28**, 569.

8 Y. D. Zhao, J. Park, R. J. Jung, H. J. Noh and S. J. Oh, *J. Mag. Mag. Mater.*, 2004, **280**, 404.

9 J. B. Goodenough, A. Wold, R. J. Arnott and N. Menyuk, *Phys. Rev.*, 1961, **124**, 373.

10 G. H. Jonker, *J. Appl. Phys.*, 1966, **37**, 1424; N. Y. Vasanthacharya, P. Ganguly, J. B. Goodenough and C. N. R. Rao, *J. Phys. C: Solid State Phys.*, 1984, **17**, 2745.

11 S. Hebert, C. Martin, A. Maignan, R. Retoux, M. Hervieu, N. Nguyen and B. Raveau, *Phys. Rev. B*, 2002, **65**, 104420.

12 P. A. Joy, Y. B. Khollam and S. K. Date, *Phys. Rev. B*, 2000, **62**, 8608.

13 R. I. Dass and J. B. Goodenough, *Phys. Rev. B*, 2003, **67**, 014401.

14  H. Z. Guo, A. Gupta, T. G. Calvarese and M. A. Subramanian, *Appl. Phys. Lett.*, 2006, **89**, 262503.

15  G. Blasse, *J. Phys. Chem. Solids*, 1965, **26**, 1969.

16  V. L. J. Joly, P. A. Joy, S. K. Date and C. S. Gopinath, *Phys. Rev. B*, 2002, **65**, 184416.

17  R. I. Dass, J. Q. Yan and J. B. Goodenough, *Phys. Rev. B*, 2003, **68**, 064415.

18  N. S. Rogado, J. Li, A. W. Sleight and M. A. Subramanian, *Adv. Mater.*, 2005, **17**, 2225.

19  C. L. Bull, D. Gleeson and K. S. Knight, *J. Phys.: Condens. Matter*, 2003, **15**, 4927.

20  M. Azuma, K. Takata, T. Saito, S. Ishiwata, Y. Shimakawa, and M. Takano, *J. Am. Chem. Soc.*, 2005, **127**, 8889.

21  M. Sakai, A. Masuno, D. Kan, M. Hashisaka, K. Takata, M. Azuma, M. Takano and Y. Shimakawa, *Appl. Phys. Lett.*, 2007, **90**, 072903.

22  H. M. Rietveld, *J. Appl. Crystallogr.*, 1969, **2**, 65.

23  N. Verelst, N. Rangavittal, C. N. R. Rao and A. Reusset, *J. Solid State Chem.*, 1993, **104**, 74; J. Topfer, and J. B. Goodenough, *J. Solid State Chem.*, 1997, **130**, 117.

24  D. N. H. Nam, R. Mathieu, P. Nordblad, N. V. Khiem and N. X. Phuc, *Phys. Rev. B*, 2000, **62**, 1027.

25  A. K. Kundu, P. Nordblad and C. N. R. Rao, *J. Phys.: Condens. Matter*, 2006, **18**, 4809.

26  K. Binder and A. P. Young, *Rev. Mod. Phys.*, 1986, **58**, 801; J. A. Mydosh, *In "Spin Glasses: An Experimental Introduction" Taylor and Francis: London*, 1993 27 A. K. Kundu, P. Nordblad and C. N. R. Rao, *J. Solid State Chem.*, 2006, **179**, 923.

**Figure captions**

**Fig. 1.** Rietveld analysis of XRD pattern for (a) $LaBiMn_2O_{6.20}$ (b) $LaBiMn_{4/3}Co_{2/3}O_{6.02}$ and (c) $LaBiMn_{4/3}Ni_{2/3}O_{5.96}$ at room temperature. Open symbols are experimental data and the dotted, solid and vertical lines represent the calculated pattern, difference curve and Bragg position respectively.

**Fig. 2.** Temperature variation of the ZFC (open symbol) and FC (solid symbol) magnetization, M, of $LaBiMn_2O_{6.20}$ at different applied fields (a) H = 100 Oe (inset shows inverse susceptibility, $\chi^{-1}$, vs temperature plot) and (b) H = 20 Oe (inset figure for H = 1000 Oe).

**Fig. 3.** (a) Field variation of magnetization at two different temperatures and (b) temperature variation in-phase component of ac-susceptibility, $\chi'$, at four different frequencies ($h_{ac}$ = 10 Oe) for $LaBiMn_2O_{6.20}$.

**Fig. 4.** Temperature variation of the ZFC (open symbol) and FC (solid symbol) magnetization, M, of $LaBiMn_{2-x}Co_xO_{6+\delta}$. The inset shows (dM/dT) vs temperature plot for FC magnetization (H = 100 Oe).

**Fig. 5.** Temperature variation of the ZFC (open symbol) and FC (solid symbol) magnetization, M, of (a) $LaBiMn_{3/2}Co_{1/2}O_{6.04}$ and (b) $LaBiMn_{4/3}Co_{2/3}O_{6.02}$ (H = 1000 Oe). The insets show typical hysteresis curves at two different temperatures.

**Fig. 6.** Temperature variation of the ZFC (open symbol) and FC (solid symbol) magnetization, M, of (a) $LaBiMn_{2-x}Ni_xO_{6+\delta}$ (H = 100 Oe) and (b) $LaBiMn_{4/3}Ni_{2/3}O_{5.96}$ (H = 1000 Oe); the inset shows typical hysteresis curve at 10 K.

**Fig. 7.** The temperature variation in-phase component of ac-susceptibility, $\chi'$, of (a) LaBiMn$_{4/3}$Co$_{2/3}$O$_{6.02}$ and (b) LaBiMn$_{4/3}$Ni$_{2/3}$O$_{5.96}$ at different frequencies ($h_{ac}$ = 10 Oe).

**Fig. 8.** Temperature variation of electrical resistivity, $\rho$, of LaBiMn$_{2-x}$(Co/Ni)$_x$O$_{6+\delta}$. The magnetoresistance effect is shown for LaBiMn$_{3/2}$Co$_{1/2}$O$_{6.04}$; $\rho$(T) is measured at 0 T (dotted line) and 7 T (solid line). The inset figure shows Seebeck coefficient, S, with the variation of temperature.

**Table. 1.** Lattice parameters, magnetic and electrical properties of LaBiMn$_{2-x}$(Co/Ni)$_x$O$_{6+\delta}$. Where $a$, $b$, $c$, R$_b$ and R$_f$ are the lattice parameters, Bragg factor and fit factor, respectively. T$_C$ is the ferromagnetic Curie temperature, θ$_p$ is the Curie-Weiss temperature, μ$_{eff}$ is the effective paramagnetic moment, M$_S$ is the approximate saturation value (10 K), H$_C$ is the coercive field (10 K), and ρ and S are the electrical resistivity and Seebeck coefficient at 300 K respectively.

| Compositions | **LaBiMn$_2$O$_{6.20}$** | **LaBiMn$_{3/2}$Co$_{1/2}$O$_{6.04}$** | **LaBiMn$_{4/3}$Co$_{2/3}$O$_{6.02}$** | **LaBiMn$_{4/3}$Ni$_{2/3}$O$_{5.96}$** |
|---|---|---|---|---|
| Space group | *Pnma* | *Pnma* | *Pnma* | *Pnma* |
| $a$ (Å) | 5.5360(8) | 5.5330(4) | 5.5292(6) | 5.5071(6) |
| $b$ (Å) | 7.8248(1) | 7.8105(6) | 7.8011(8) | 7.7790(1) |
| $c$ (Å) | 5.5455(7) | 5.5200(4) | 5.5130(5) | 5.5009(9) |
| Cell volume (Å$^3$) | 240.22(6) | 238.54(3) | 237.8(4) | 235.66(6) |
| R$_b$ | 8.33 | 5.60 | 5.35 | 5.44 |
| R$_f$ (%) | 9.89 | 8.47 | 9.07 | 8.36 |
| δ in LaBiMn$_{2-x}$M$_x$O$_{6+\delta}$ | 0.20 | 0.04 | 0.02 | -0.04 |
| Average valency for (Mn,Co/Ni) | 3.20 | 3.04 | 3.02 | 2.96 |
| Average valency for Mn if M$^{2+}$ | 3.20 | 3.39 | 3.53 | 3.44 |
| T$_C$ (K) | 80 | 87 | 130 | 97 |
| θ$_p$ (K) | 120 | 135 | 150 | 140 |
| μ$_{eff}$ (μ$_B$/f.u.) | 7.88 | 9.05 | 7.10 | 6.03 |
| M$_S$ (μ$_B$/f.u.) | 6.01 | 7.72 | 4.89 | 3.47 |
| H$_C$ (Tesla) | 0.047 | 0.181 | 0.575 | 0.003 |
| ρ$^{300K}$ (Ω.cm) | 26 | 14 | 43 | 33 |
| S$^{300K}$ (μV/K) | 133 | - | 81 | - |

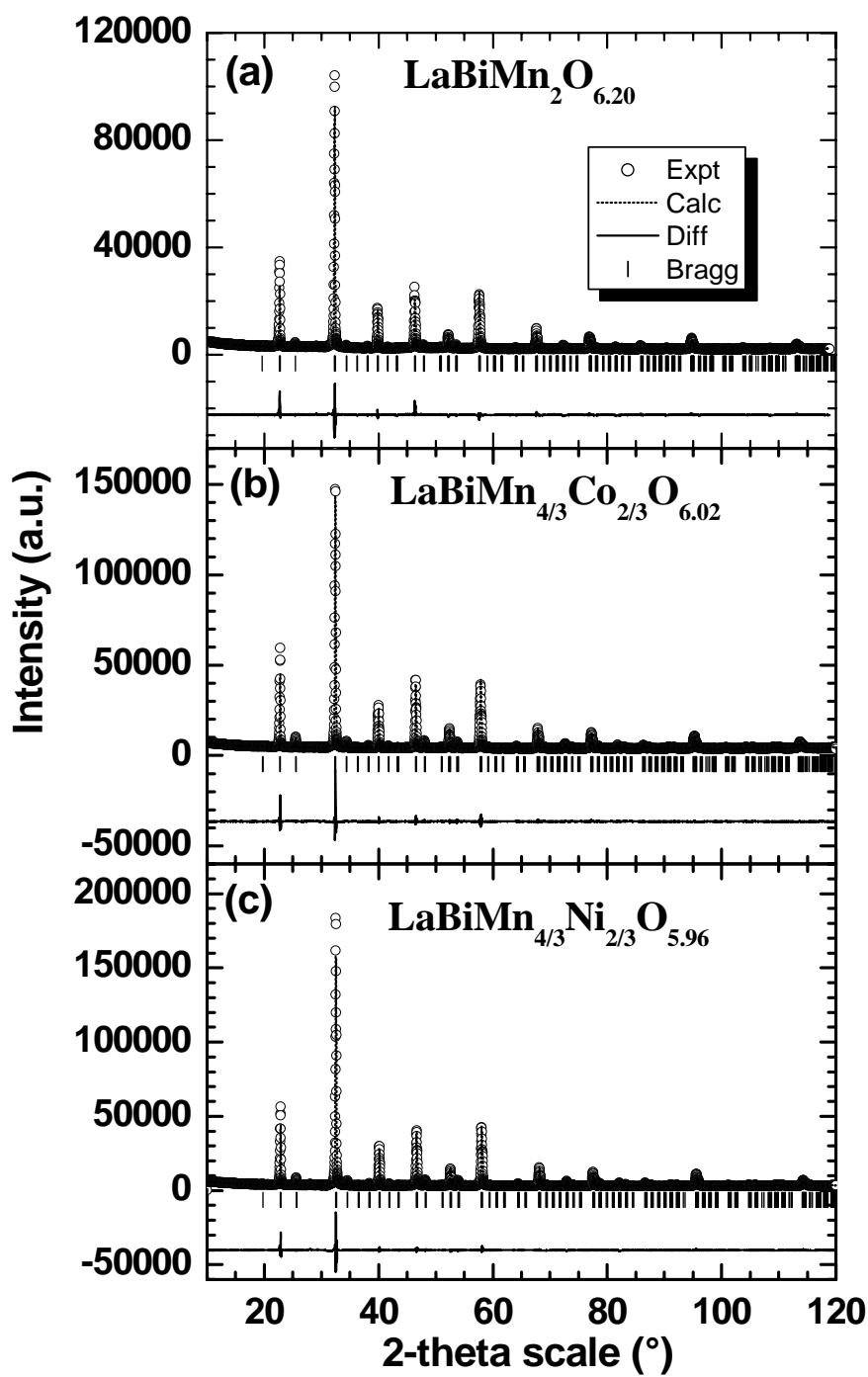

**Fig. 1.** Rietveld analysis of XRD pattern for (a) $LaBiMn_2O_{6.20}$ (b) $LaBiMn_{4/3}Co_{2/3}O_{6.02}$ and (c) $LaBiMn_{4/3}Ni_{2/3}O_{5.96}$ at room temperature. Open symbols are experimental data and the dotted, solid and vertical lines represent the calculated pattern, difference curve and Bragg position respectively.

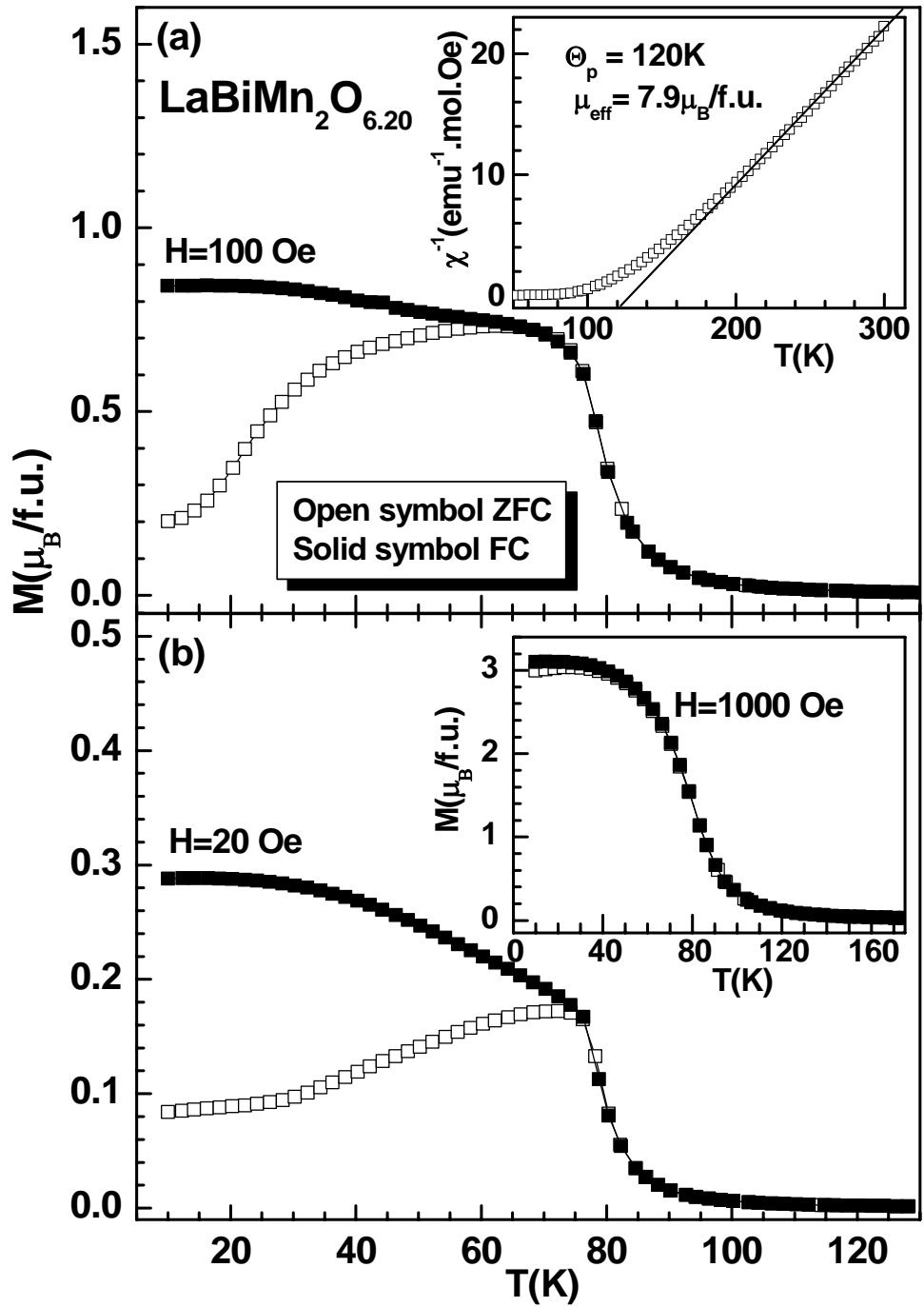

**Fig. 2.** Temperature variation of the ZFC (open symbol) and FC (solid symbol) magnetization, M, of LaBiMn$_2$O$_{6.20}$ at different applied fields (a) H = 100 Oe (inset shows inverse susceptibility, $\chi^{-1}$, vs temperature plot) and (b) H = 20 Oe (inset figure for H = 1000 Oe).

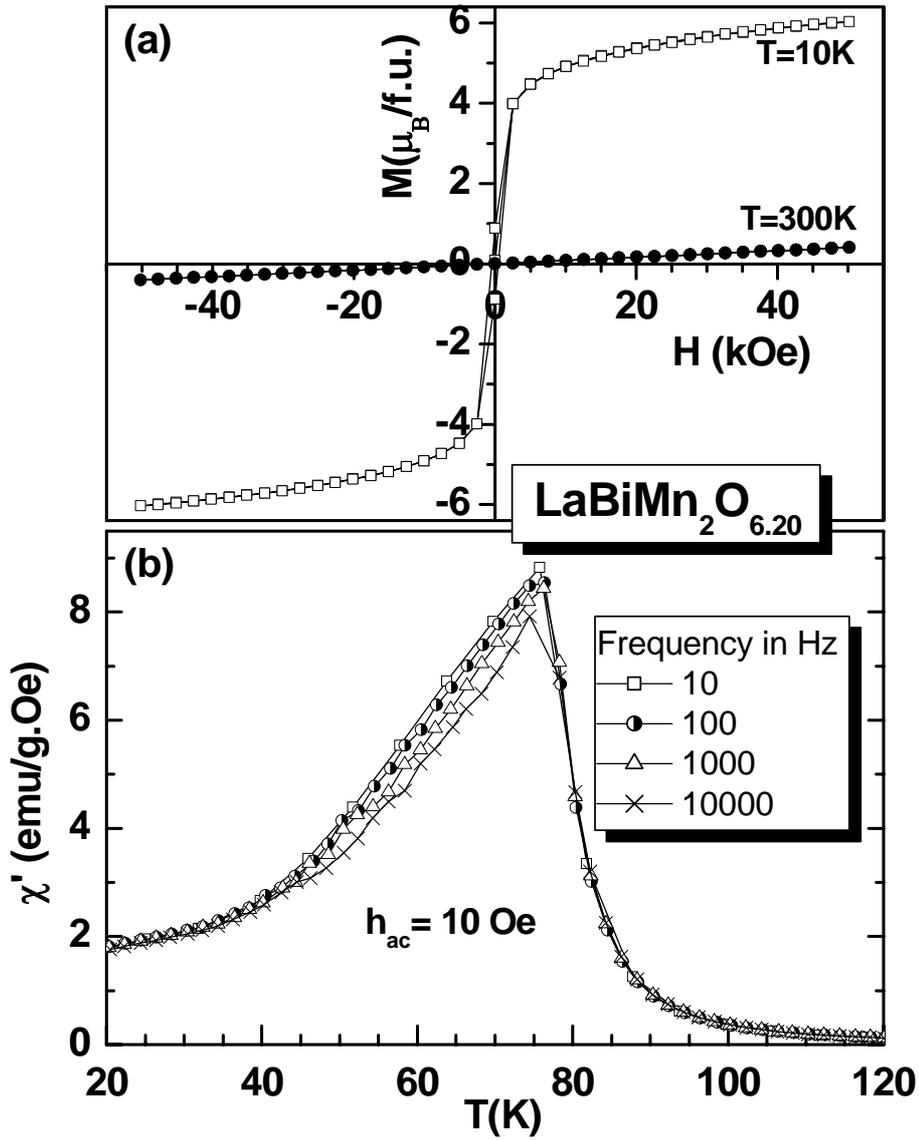

**Fig. 3.** (a) Field variation of magnetization at two different temperatures and (b) temperature variation in-phase component of ac-susceptibility, $\chi'$, at four different frequencies ($h_{ac}$ = 10 Oe) for $LaBiMn_2O_{6.20}$.

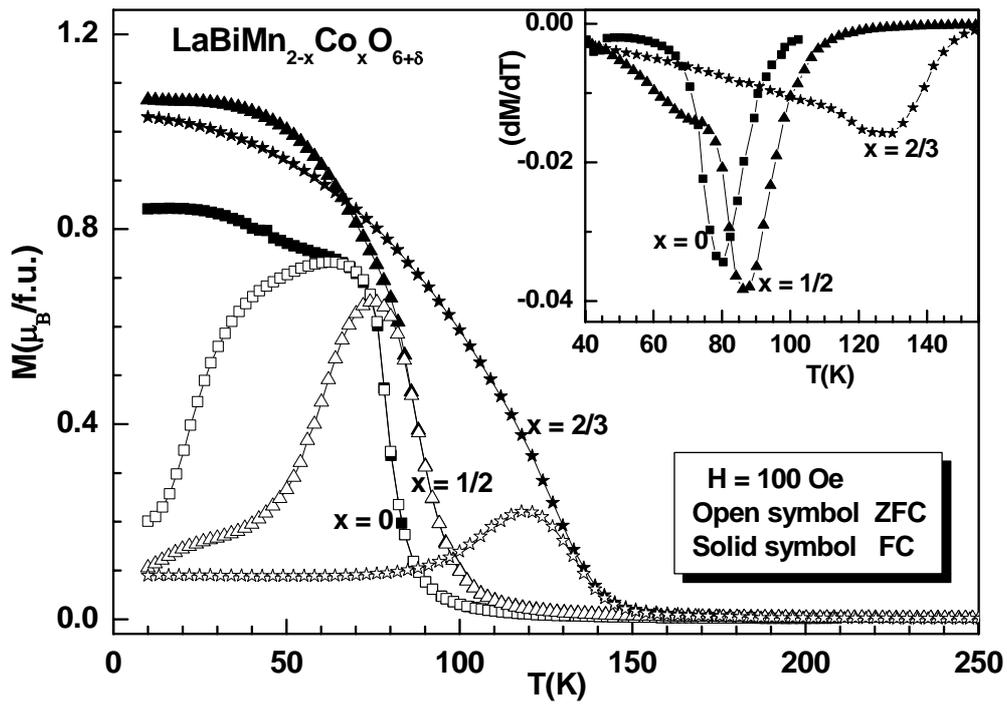

**Fig. 4.** Temperature variation of the ZFC (open symbol) and FC (solid symbol) magnetization, M, of $LaBiMn_{2-x}Co_xO_{6+\delta}$. The inset shows (dM/dT) versus temperature plot for FC magnetization (H = 100 Oe).

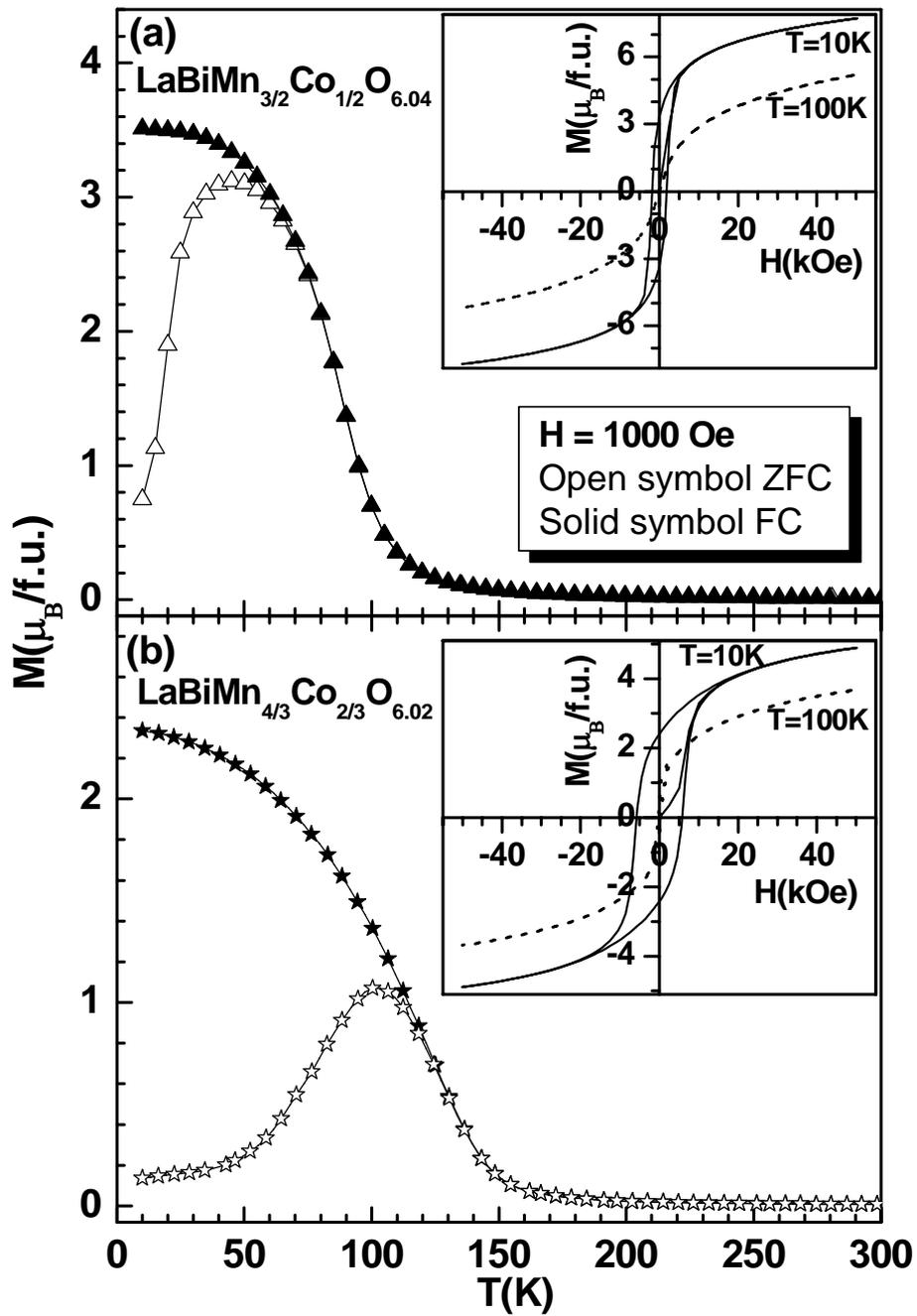

**Fig. 5.** Temperature variation of the ZFC (open symbol) and FC (solid symbol) magnetization, M, of (a) LaBiMn$_{3/2}$Co$_{1/2}$O$_{6.04}$ and (b) LaBiMn$_{4/3}$Co$_{2/3}$O$_{6.02}$ (H = 1000 Oe). The insets show typical hysteresis curves at two different temperatures.

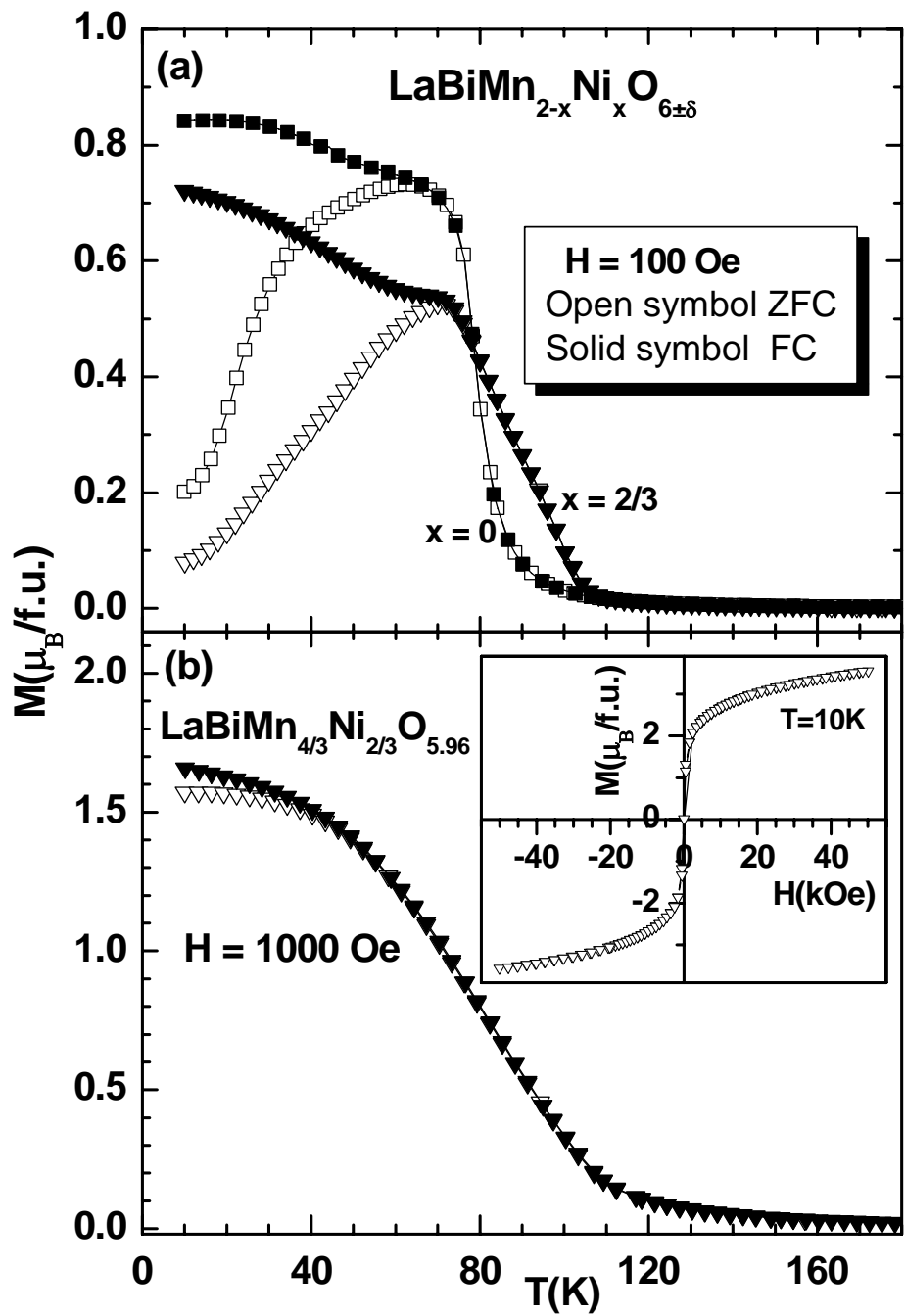

**Fig. 6.** Temperature variation of the ZFC (open symbol) and FC (solid symbol) magnetization, M, of (a) $LaBiMn_{2-x}Ni_xO_{6+\delta}$ (H = 100 Oe) and (b) $LaBiMn_{4/3}Ni_{2/3}O_{5.96}$ (H = 1000 Oe); the inset shows typical hysteresis curve at 10 K.

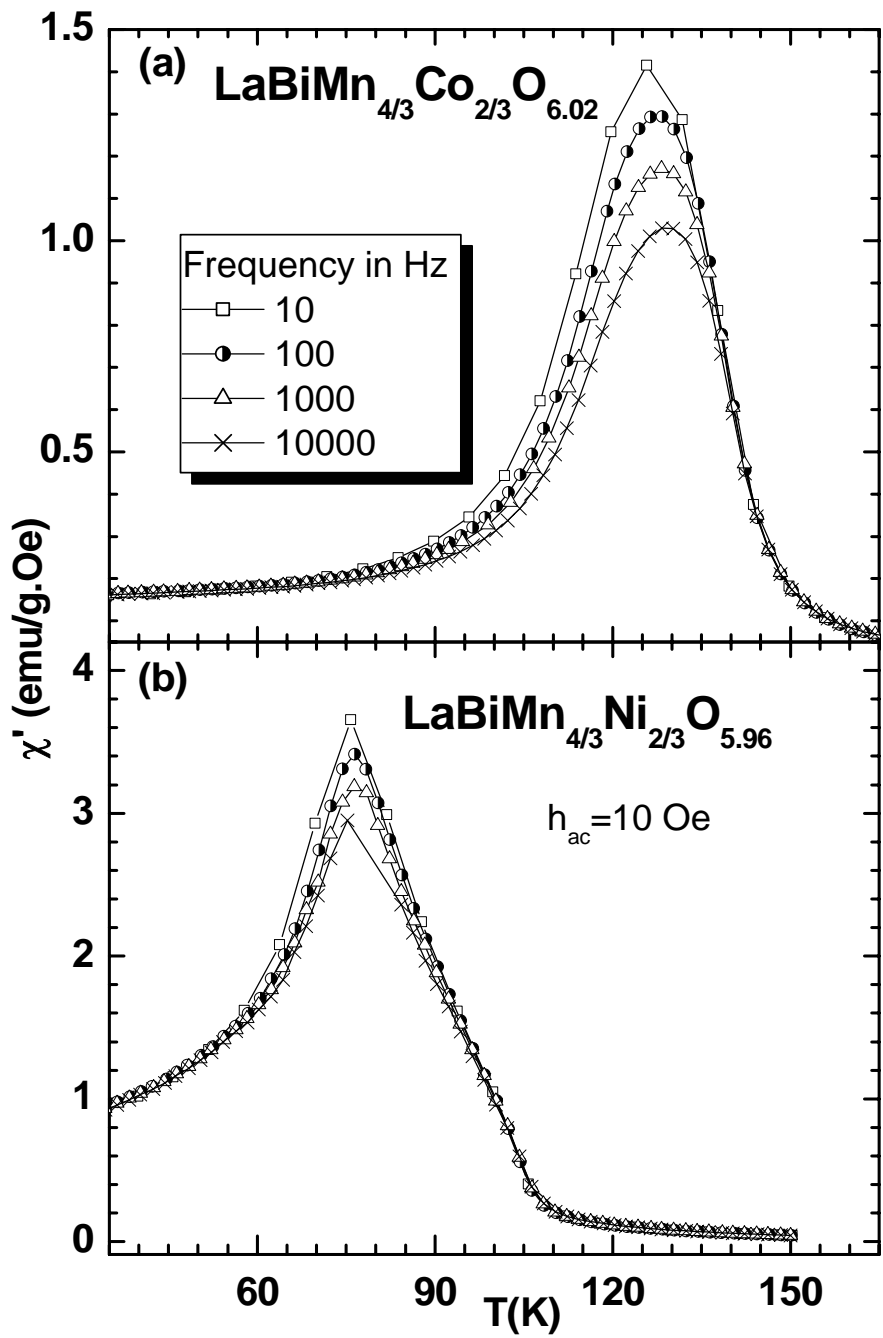

**Fig. 7.** The temperature variation in-phase component of ac-susceptibility, $\chi'$, of (a) $LaBiMn_{4/3}Co_{2/3}O_{6.02}$ and (b) $LaBiMn_{4/3}Ni_{2/3}O_{5.96}$ at different frequencies ($h_{ac}$ = 10 Oe).

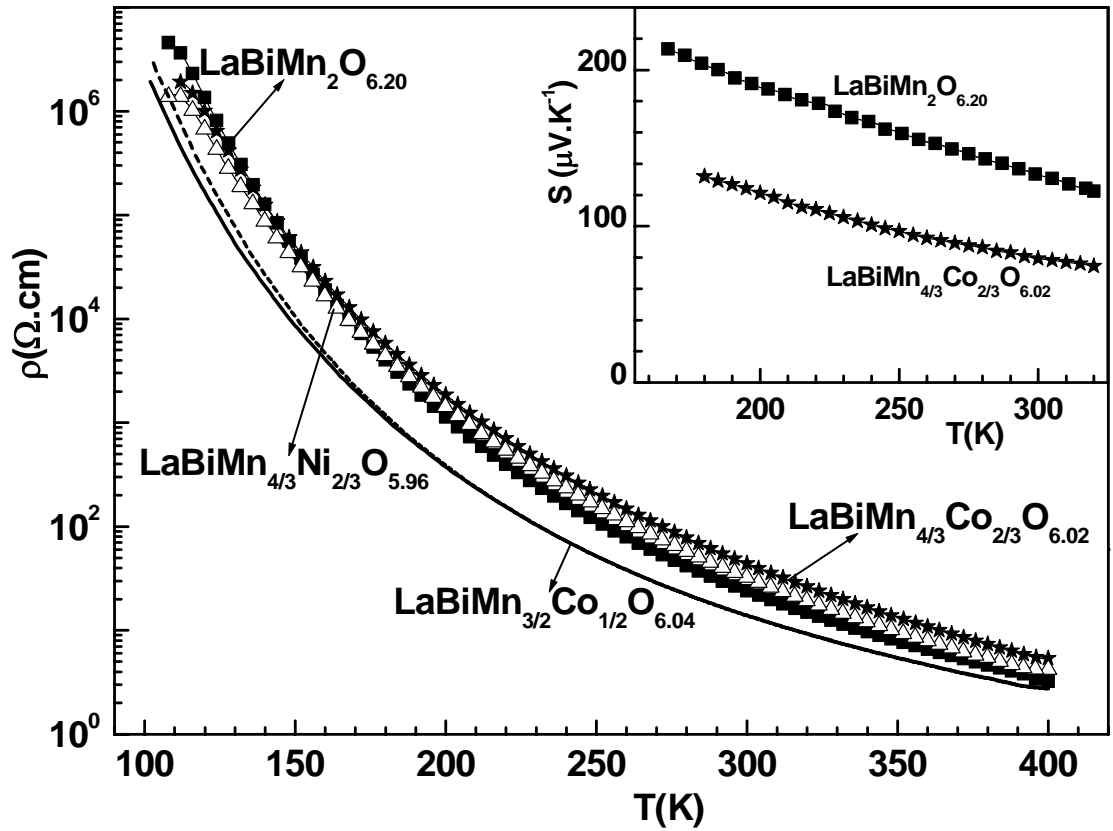

**Fig. 8.** Temperature variation of electrical resistivity, ρ, of $LaBiMn_{2-x}(Co/Ni)_xO_{6+\delta}$. The magnetoresistance effect is shown for $LaBiMn_{3/2}Co_{1/2}O_{6.04}$; ρ(T) is measured at 0 T (dotted line) and 7 T (solid line). The inset figure shows Seebeck coefficient, S, with the variation of temperature.

**Graphical abstract:**

Temperature dependence ZFC (open symbol) and FC (solid symbol) magnetization, M, of $LaBiMn_{2-x}Co_xO_{6+\delta}$.

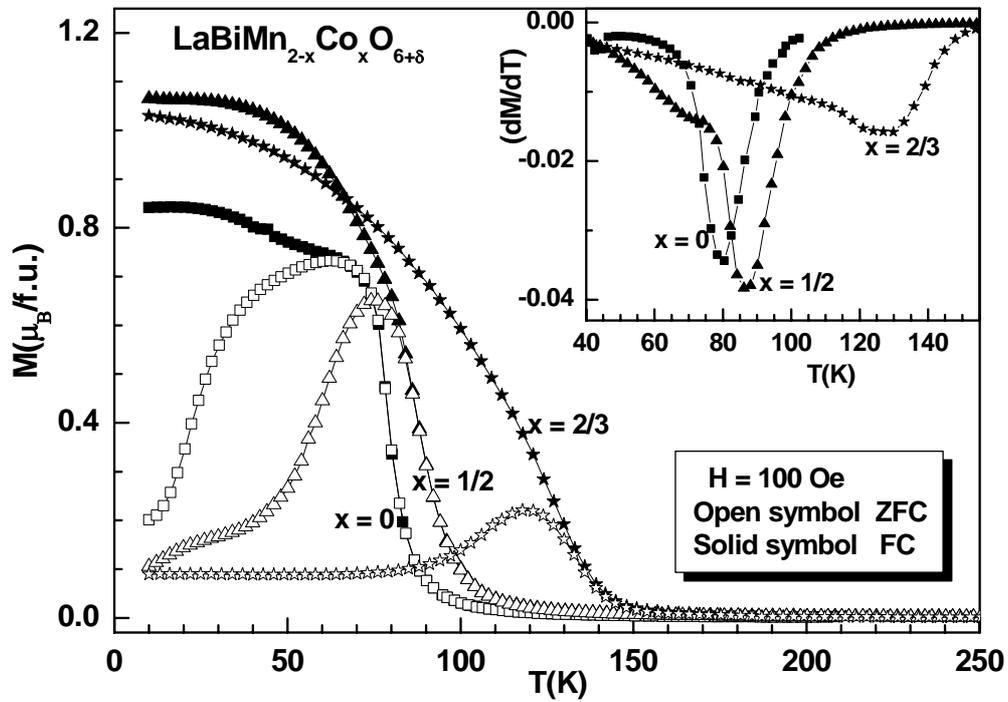